\documentclass[a4paper,12pt]{article}
\usepackage{jheppub} 
\usepackage{lineno}
\usepackage{amsmath}
\usepackage{amsfonts}
\usepackage{amssymb}
\usepackage{graphicx}
\usepackage{lmodern}
\usepackage{dcolumn}
\usepackage{bm}
\usepackage{float}
\usepackage[T1]{fontenc}
\usepackage{subfigure}
\usepackage{hyperref}
\usepackage{multirow}
\usepackage{array}
\DeclareUnicodeCharacter{2064}{\ensuremath{-}}
\DeclareUnicodeCharacter{2212}{\ensuremath{-}}

\title{\boldmath Unveiling Neutrino Mysteries with $\Delta(27)$ Symmetry.}

\author[a]{Manash Dey}
\author[a]{Subhankar Roy}
\affiliation[a]{Department of Physics, Gauhati University, India, 781014}

\emailAdd{manashdey@gauhati.ac.in}
\emailAdd{subhankar@gauhati.ac.in}
\abstract{An elegant model is proposed by extending the Standard Model using the $\Delta(27)\times Z_3 \times Z_{10}$ symmetry within the framework of the Type-I + Type-II seesaw mechanism. This model is particularly noteworthy for its ability to restrict the atmospheric mixing angle, $\theta_{23}$, to specific values, and provides an explanation for the observed hierarchy of charged lepton masses. The neutrino mass matrix texture defined by three real parameters, predicts the three neutrino mass eigenvalues and the two Majorana phases. Furthermore, the model is tested against the experimental results of neutrino-less double beta ($0\nu\beta\beta$) decay and charged lepton flavour violation (cLFV) experiments.}

\begin{document}
\maketitle
\flushbottom

\section{Introduction}
\label{sec:intro}
Over the past few decades, the Standard Model (SM) of particle physics, characterized by the gauge group \(SU(2)_L \times U(1)_Y\),  has gained widespread acceptance as a comprehensive theoretical framework that encompasses the fundamental particles, such as quarks and leptons and explains the fundamental interactions viz., the strong, weak, and electromagnetic interactions. Even though the SM is very successful in explaining a wide range of phenomena,  but the theory is inefficient to explain neutrino mass origin, neutrino mass hierarchy, gravity, matter-antimatter asymmetry etc. Out of the many shortcomings, the inability to address the issues related to neutrino masses and mixing are the captivating problems in the realm of particle physics. The neutrinos~\cite{Cowan:1956rrn} are massless in the framework of the SM. However, in the year 1957, Pontecorvo anticipated that neutrinos could change their flavour as they travel in space, a phenomenon known as neutrino oscillation~\cite{Pontecorvo:1957cp}. The latter after experimental verification~\cite{ SNO:2002tuh, KamLAND:2002uet, Super-Kamiokande:1998uiq}, in the recent past confirmed the fact that neutrinos are not massless. Therefore, it is apt to assert that the pursuit of finding answers to the missing pieces in the SM compels the model builders to go beyond the latter (BSM). In the context of the BSM theories, the neutrino mass matrix is an important quantity. The neutrino mass matrix, $M_{\nu}$, originates from the Yukawa Lagrangian\,($\mathcal{L}_Y$), and it contains the information of masses and mixing. It encodes the information of all observational parameters which includes the solar mixing angle ($\theta_{12}$), reactor mixing angle ($\theta_{13}$), atmospheric mixing angle ($\theta_{23}$), the neutrino mass eigenvalues ($m_1, m_2, m_3$), a Dirac CP-violating phase ($\delta$), as well as two Majorana phases, namely $\alpha$ and $\beta$. The neutrino mass matrix is formulated within the framework of seesaw mechanism~\cite{Yoshimura:1978ex, Akhmedov:1999tm}.

	   There are many forms of seesaw mechanisms proposed in the literature. For example, the Type-I seesaw mechanism~\cite{Cai:2017mow, Mohapatra:2004zh, King:2013eh, Mohapatra:2006gs, King:2003jb} is a simple extension of the SM, it introduces right-handed neutrinos ($\nu_R$) to create both Majorana and Dirac mass terms, represented as $M_R (\overline{\nu^c_R}\nu_{R})$ and $M_D (\bar{\nu}_L \nu_R)$ respectively. Here, the $\nu_L$ denotes the left-handed neutrino field from the SM doublet, $M_D$ is the Dirac neutrino mass matrix and $M_R$ is the right handed neutrino mass matrix. If one assumes that the scale of $M_R$ is much larger than $M_D$, the neutrino mass matrix in Type-I seesaw is given by the expression,
\begin{equation}
-M_D. M^{-1}_R. M^T_D,
\end{equation}	   
and the neutrino masses are observed to be of the order of around $10^{-2}$ eV. Therefore, the smallness of neutrino masses in this scenario can be attributed to the significant scale of $M_R$, which functions as a seesaw mechanism.

		There is another type of seesaw mechanism known as Type-II seesaw ~\cite{Melfo:2011nx, FileviezPerez:2008jbu, King:2003jb, Cheng:1980qt, Cai:2017mow}. Here, one introduces a heavy $SU(2)_L$ scalar triplet, $\Delta$, in the Higgs sector of the SM. The light Majorana neutrino mass term  in this scenario, is given by $\sim y_{\Delta}(\bar{D}_{l_L}\, D_{l_L}^c)\,i\sigma_2 {\Delta} $. When $m_{\Delta} \gg v_{h}$, the neutral component of the scalar triplet acquires a non-zero vacuum expectation value (vev), $v_{\Delta}$, it leads to a small Majorana neutrino mass: $m_\nu\sim y_{\Delta}v_{\Delta}$, where, $v_{\Delta} \sim v_{h}^2/m_{\Delta}^2$ and $y_{\Delta}$ is the corresponding coupling constant. Here, $m_{\Delta}$ represents the mass of the scalar triplet, and $v_{h}$ signifies the SM Higgs vev ($v_h\sim$ 246 GeV). This is to be underlined that one can generate the effective Majorana neutrino mass matrix in a framework where both Type-I and Type-II seesaw mechanisms may coexist. Such frameworks, most often, are referred to as the Type-I + Type-II seesaw mechanism ~\cite{Cai:2017mow, Akhmedov:2006de, Wong:2022qyg, Kashav:2023tmz, Banerjee:2023aro}. 

	Though the neutrino mass matrix\,($M_{\nu}$) carries the information of all the observational parameters, but from the neutrino oscillation experiments, we can estimate only the mixing angles and $\delta$. The neutrino oscillation experiments cannot give the exact values of the neutrino mass eigenvalues and the two Majorana phases. In this regard, if we can workout certain correlations among the neutrino mass matrix elements then we may address the problem of neutrino masses and mixing. A neutrino mass matrix exhibiting correlations are referred to as ``textures''. In this connection, various phenomenological ideas have been proposed. These ideas include concepts like texture zeroes ~\cite{Ramond:1993kv, Ibarra:2003Jb, Ludl:2014axa}, $\mu$-$\tau$ symmetry ~\cite{Harrison:2002er, Ma:2005qf, Lin:2009bw, Xing:2020ijf}, $\mu-\tau$ mixed symmetry ~\cite{Dey:2022qpu}, $\mu-\tau$ reflection symmetry ~\cite{Xing:2022uax, Liu:2017frs}, $\mu-\tau$ antisymmetry ~\cite{Xing:2015fdg, Xing:2020ijf}, and more. These ideas are attributed to the neutrino mass matrix, $M_{\nu}$, to place constraints on the neutrino mass matrix elements and to draw significant phenomenological insights.

	In this light, we shall try to formulate a similar texture starting from 
	the seesaw mechanism, which predicts significantly on the oscillation parameters. The present work is interesting, as  it shows a forbidden region for the atmospheric mixing angle, $\theta_{23}$.

		The structure of the work is as follows: In Section ~\ref{section 2}, we introduce the model, with a brief discussion of the discrete flavour symmetry, $\Delta(27)$ ~\cite{Branco:1983tn}, which is employed in developing the neutrino mass matrix. We then delve into the field content, Lagrangian, and scalar potential of our model. In Section ~\ref{section 3}, we discuss the numerical analysis and key findings of our work. In Section ~\ref{section 4} and Section ~\ref{section 5}, we test the predictability of the model in light of Neutrinoless Double Beta\,($0\nu\beta\beta$) Decay and Charged Lepton Flavour Violation\,(cLFV) respectively. Finally, in Section\, \ref{section 6}, we highlight the summary and discussions of our work.
\section{Theoretical Framework}
\label{section 2}
We extend the SM gauge group with $\Delta(27)$ symmetry~\cite{Branco:1983tn, Ma:2007wu, Abbas:2014ewa, Chen:2015jta, CentellesChulia:2016fxr, Vien:2020hzy}. The discrete symmetry group $\Delta(27)$ has eleven irreducible representations, out of which there are one triplet\,($3$), one anti-triplet \,($\bar{3}$) and nine singlet\,($1_{p,q}$) representations, where, $p,q = 0, 1, 2$. Unlike simple discrete symmetry groups such as $A_4$\,\cite{Ma:2001dn, King:2006np, Altarelli:2010gt, King:2013eh}, which has got only two triplets ($3$) and three singlet ($1, 1^{'}, 1^{''}$) representations, the presence of  an additional anti-triplet ($\bar{3}$) representation within the framework of $\Delta(27)$ symmetry provides a greater flexibility for extracting important phenomenological insights.
	
		We modify the SM field content by adding three singlet heavy right handed neutrinos viz., $\nu_{l_R(l= e, \mu, \tau)}$, which under $\Delta(27)$ transform as $1_{00}, 1_{01}\,\text{and}\, 1_{02}$. Two additional $\Delta(27)$ triplet scalar fields, $\chi\, \text{and} \,\Delta$ are introduced, which individually transform as $1$ and $3$ under $SU(2)_L$. In addition to these, we consider five $SU(2)_L$ singlets  $\eta, \kappa, \xi$, $\zeta$ and $\varrho$, transforming as $1_{01}, 1_{02}, 1_{00}, 1_{00}$ and $1_{00}$  under $\Delta(27)$ respectively. Two additional symmetries, $Z_3$\,\cite{Ma:2004yx,Hu:2006wk} and $Z_{10}$\,\cite{CarcamoHernandez:2020udg, Dey:2023rht, Dey:2024ctx} are introduced to restrict some undesirable and next to leading order terms in the Yukawa Lagrangian. We discuss the multiplication rules of $\Delta(27)$ and $Z_{10}$ symmetries in the Appendix\,\ref{Appendix}.

	 The transformation of all the field contents under $SU(2)_L \times \Delta(27) \times Z_3 \times Z_{10}$ symmetry is highlighted in the Table\,(\ref{table:1}).
\begin{table}[h!]
\centering
\begin{tabular}
{p{1.5cm}p{0.4cm}p{0.4cm}p{1.6cm}p{0.4cm}p{2cm}p{0.3cm}p{0.3cm}p{0.3cm}p{0.3cm}p{0.3cm}p{0.3cm}p{0.3cm}}
\hline
\noalign{\vskip 0.8mm}
Fields & $\bar{D}_{l_{L}}$ & $D_{l_L}^c$ & \quad\,\,$l_R$ & $H$ &\quad \,\, $\nu_{l_R}$ &$\chi$ &  $\Delta$ & $\eta$ &$\kappa$&$\xi$& $\zeta$&$\varrho$ \\
\hline
$SU(2)_{L}$ & 2 & 2 & \quad\,\, 1 & 2 &\quad \quad 1 & 1 & 3 & 1 & 1 &1& 1 & 1 \\
\hline
\noalign{\vskip 0.8mm}
$\Delta(27)$ & 3 & 3 & \quad\,\, $1_{0r}$ & $1_{00}$ &\quad \,\, $1_{0r}$ & $\bar{3}$ & 3 & $1_{01}$&$1_{02}$&$1_{00}$&$1_{00}$&$1_{00}$ \\
\hline
$Z_{3}$ & 1 & 1 &\quad\,\, 1 & 1 & (1, $\omega^*$, $\omega^*$) & 1 & 1 & $\omega^*$&$\omega^*$&$\omega$&$\omega$&$\omega^*$ \\
\hline
$Z_{10}$ & 0 & 0 &\quad(1,4,7) & 0 & \quad(0,2,7) & 0 & 0 & 6&6&8&3&1 \\
\hline
\end{tabular}
\caption{ The transformation properties of the field contents under $SU(2)_{L} \times \Delta(27)\times Z_3 \times Z_{10}$. Where, $r= 0,1,2$, $l= e, \mu, \tau$ and $\omega= e^{2 \pi i/3}$.} 
\label{table:1}
\end{table}

	The relevant Yukawa Lagrangian\,($ \mathcal{L}_Y$) of the model is shown below,
\begin{eqnarray}
- \mathcal{L}_Y &=& \mathcal{L}_{l}+ \mathcal{L}_{\nu}.
\end{eqnarray}
Where, $\mathcal{L}_{l}$ and $\mathcal{L}_{\nu}$, are for the charged leptons and neutrinos respectively, and they are presented in the following manner,
\begin{eqnarray}
\mathcal{L}_{l} &=& \frac{y_1}{\Lambda^{10}} (\bar{D}_{l_L}\chi)\,H\, e_{R} \varrho^9 +\,\frac{y_2}{\Lambda^7} (\bar{D}_{l_L}\chi)\,H\,\mu_{R}\varrho^6+\frac{y_3}{\Lambda^4} (\bar{D}_{l_L}\chi)\,H \,\tau_{R}\varrho^3+ \, h.c.\\
\label{clagrangian}
\nonumber\\
\text{and}\nonumber\\
\nonumber\\
\mathcal{L}_{\nu}&=& \frac{y_e}{\Lambda}\,(\bar{D}_{l_L}\chi)\,\tilde{H}\,\nu_{e_R} + \frac{y_\mu}{\Lambda^2}\,(\bar{D}_{l_L}\chi)\,\tilde{H}\,\nu_{\mu_{R}}\,\xi  +\frac{y_\tau}{\Lambda^2}\,(\bar{D}_{l_L}\chi)\,\tilde{H}\,\nu_{\tau_{R}}\,\zeta +\,\frac{1}{2}\,M_1\,\overline{\nu^c_{e_R}}\,\nu_{e_R} +\,\frac{1}{2}\nonumber\\&&\,y_{s}\,[\overline{\nu^c_{\mu_R}}\nu_{\tau_{R}}+\, \overline{\nu^c_{\tau_R}}\,\nu_{\mu_{R}}]\varrho+\,\frac{1}{2}\,y_{r_1}(\overline{\nu^c_{\mu_R}}\,\nu_{\mu_{R}})\eta +\,\,\frac{1}{2}\,y_{r_2}\,(\overline{\nu^c_{\tau_R}}\,\nu_{\tau_{R}})\,{\kappa} +\, y_{\Delta}(\bar{D}_{l_L}\, D_{l_L}^c)\,\tilde{\Delta}\nonumber\\&&+ h.c.
\label{nLagrangian}
\end{eqnarray} 
Here, $\tilde{H}= i \sigma_2 H^*$, $\tilde{\Delta}= i \sigma_2 \Delta$ and $\Lambda$ is the effective scale of the theory. After the scalars develop their respective vevs in the following way: $\langle \chi \rangle = v_{\chi}(1,0,0)$, $\langle H \rangle = v_h, \langle \Delta \rangle = v_{\Delta}(1,1,1), \langle \eta \rangle = v_{\eta}, \langle \kappa \rangle = v_{\kappa}, \langle \varrho \rangle = v_{\varrho}, \langle \xi \rangle = v_{\xi}$ and  $\langle \zeta \rangle = v_{\zeta}$, the charged leptons and the neutrinos acquire their masses. In this context, we wish to highlight that $v_h$, $v_\Delta$ and $v_\chi$ exhibit the following relationship,
\begin{equation}
v_{\Delta} v_{\chi} \sim \frac{2 }{3}v_h^2.
\end{equation}
This relation arises from the scalar potential, as discussed in the Appendix \,\ref{Appendix}.
From this relation, we can estimate the vev, $v_{\chi}$. By choosing $v_{\Delta}$ in the range of approximately $\sim$ $1$eV - $1$ GeV\,\cite{FileviezPerez:2008jbu}, and considering that $v_h= 246$ GeV, we find that $v_{\chi}$ lies in the range of approximately $\sim$ $4$ TeV - $7 \times 10^6$ TeV.

Furthermore, it is important to mention that the $Z_3$ symmetry of $ \mathcal{L}_Y$ is broken by the vev of the scalar singlet $\varrho$. This breaking is important for the explanation of the charged lepton mass hierarchy \,\cite{Froggatt:1978nt,Babu:1999me,Lykken:2008bw,Hagedorn:2010mq,Ganguly:2022cbo,Bonilla:2023wok,CarcamoHernandez:2024vcr}.

	From the $\mathcal{L}_{l}$, we extract the information of the charged lepton mass matrix ($M_L$), and it takes the form,
\begin{eqnarray}
M_{L}&=& v_h v_{\chi} \begin{bmatrix}
\frac{v_{\varrho}^9 y_1}{\Lambda^{10}} & 0 & 0\\
0   & \frac{v_{\varrho}^6 y_2}{\Lambda^{7}} & 0\\
0   &  0     & \frac{v_{\varrho}^3 y_3}{\Lambda^4}\\
\end{bmatrix}.
\end{eqnarray}
The matrix $M_L$ is a diagonal matrix, and it is observed that the charged lepton masses satisfy the following relation,
\begin{equation}
m_e \colon m_{\mu} \colon m_{\tau} \approx \lambda^6_m \, y_1 \colon \lambda^3_m \,y_2 \colon y_3,
\label{clmh}
\end{equation}
where, $\lambda_m= v_{\varrho}/\Lambda$. The above relation is in good agreement with the observed hierarchy of the charged lepton masses. The observed hierarchy is\,\cite{ParticleDataGroup:2018ovx, ParticleDataGroup:2022pth}, 
\begin{equation}
m_e \colon m_{\mu} \colon m_{\tau} \approx 0.00028 \colon 0.05946 \colon 1.
\label{oclmh}
\end{equation}
To have a numerical understanding of the hierarchy predicted from our model,  we may take the Yukawa couplings as, $y_1 \approx 0.1$, $y_{2} \approx 1$ and $y_{3} \approx 1$, and $\lambda_m \approx 0.38$. We find that, for these inputs, Eq.\,(\ref{clmh}) is consistent with the observed hierarchy highlighted in Eq.\,(\ref{oclmh}).

To extract the information of the neutrino masses, we rewrite the $\mathcal{L}_{\nu}$ in the following way,
\begin{eqnarray}
\mathcal{L}_{\nu}&=& \frac{1}{2}\bar{\nu_L} M_{II}\nu^c_L + \bar{\nu_L}M_D \nu_R + \frac{1}{2}\overline{\nu^c_R} M_{R}\nu_R + h.c.\nonumber\\
\label{NLagrangian}
\end{eqnarray}
where, $M_D$ is the Dirac neutrino mass matrix, $M_R$ is the right handed Majorana neutrino mass matrix and $M_{II}$ is the mass matrix arising from Type-II seesaw mechanism. Although we have not written the flavour indices, it is understandable that $M_D$, $M_R$ and $M_{II}$ are $3\times 3$ matrices in the flavour space. They are shown below,
\begin{eqnarray}
M_D &=& \begin{bmatrix}
\frac{v_h v_{\chi}y_e}{\Lambda} & 0 & 0\\
0   & \frac{v_h v_{\chi} v_{\xi}y_{\mu}}{\Lambda^2} & 0\\
0   &  0     & \frac{v_h v_{\chi} v_{\zeta}y_{\tau}}{\Lambda^2}\\
\end{bmatrix},\quad
M_R = \begin{bmatrix}
 M_1 & 0 & 0\\
 0 & y_{r_1}v_{\eta} & y_s \varrho \\
 0 & y_s \varrho  & y_{r_2}v_{\kappa}  \\
 \end{bmatrix},\quad
 M_{II} = \frac{y_{\Delta}v_{\Delta}}{2}\begin{bmatrix}
  2&1&1\\
  1&2&1 \\
 1&1&2 \\
 \end{bmatrix}.\nonumber\\
 \nonumber\\
\end{eqnarray}
To understand the effective light neutrino masses, we can rewrite Eq.\,(\ref{NLagrangian}) in the following way, 
\begin{eqnarray}
\mathcal{L}_{\nu}&=& \frac{1}{2}\overline{n^c} M n + h.c,
\label{NNLagrangian}
\end{eqnarray}
with 
\begin{eqnarray}
n &=& \begin{bmatrix}
\nu^c_L\\
\nu_R\\
\end{bmatrix},
\end{eqnarray}
and the $6\times6$ mass matrix,
\begin{eqnarray}
M &=& \begin{bmatrix}
M_{II} & M_D\\
M^T_{D}   & M_R\\
\end{bmatrix}.
\end{eqnarray}
The block diagonalisation of $M$ gives the light effective neutrino mass matrix,
\begin{equation}
M_{\nu} = M_{II}-M_{D}. M_{R}^{-1}.M_{D}^{T}
\end{equation}
Here, $M_{I}= -M_{D}. M_{R}^{-1}.M_{D}^{T}$ is the contribution arising from Type-I seesaw mechanism. The complex  Majorana neutrino mass matrix\,($M_{\nu}$) in the Type-I+Type-II seesaw framework after simplification, takes the following form,
\begin{equation}
 M_{\nu} = 
 \begin{bmatrix}
 t-d  & t/2 & t/2 \\
t/2& t+a & \frac{t}{2}-c\\
t/2& \frac{t}{2}-c  & t+b \\
 \end{bmatrix},
 \label{Mass matrix}
\end{equation} 
such that the \emph{model parameters} viz., $y_e, y_{\mu}$,
 $y_{\tau}, y_{r_1}, y_{r_2}, y_s, y_{\Delta}, v_{\chi}, v_{\Delta}, v_{h}, v_{\eta}, v_{\kappa}, v_{\varrho}, v_{\xi}$, $v_{\zeta}$ and $M_1$ are related to the \emph{texture parameters} viz., $a, b, c, d $ and $t$ in the following way,
 \begin{eqnarray}
 \label{Model Parameters1}
  a &=& \frac{y_{r_2} v_{\kappa} v_h^2 v_{\xi }^2 v_{\chi }^2 y_{\mu }^2}{\Lambda ^4 \left((y_s v_{\varrho})^2-y_{r_1} y{r_2} v_\eta v_{\kappa}\right)},\\
 b &=& \frac{y_{r_1} v_{\eta} v_h^2 v_{\zeta }^2 v_{\chi }^2 y_{\tau }^2}{\Lambda ^4 \left((y_s v_{\varrho})^2-y_{r_1} y{r_2} v_\eta v_{\kappa}\right)},\\
 c &=& \frac{y_{s} v_{\varrho} v_h^2 v_\xi v_{\zeta } v_{\chi }^2 y_\mu y_{\tau }}{\Lambda ^4 \left((y_s v_{\varrho})^2-y_{r_1} y{r_2} v_\eta v_{\kappa}\right)},\\
 d &=& \frac{y_e^2 v_h^2 v_{\chi }^2}{\Lambda ^2 M_1},\\
t&=& y_{\Delta }v_{\Delta } .
 \label{Model Parameters2}
 \end{eqnarray}
 The texture contains a correlation: $(M_{\nu})_{12}=(M_{\nu})_{13}$, and it is observed that the $\mu-\tau$ symmetry is broken as $a \neq b$.
	
		Now, in order to draw predictions from a mass matrix, one needs to diagonalise it with the help of a diagonalising matrix. In the context of neutrino physics phenomenology, this diagonalising matrix is known as the Pontecorvo-Maki-Nakagawa-Sakata(PMNS) matrix, $\tilde{U}$ ~\cite{Maki:1962mu}. A detailed discussion on the parametrisation of the PMNS matrix is shown in the Appendix\,\ref{Appendix}.

		The diagonalising matrix $\tilde{U}$, constrains the $M_{\nu}$ such that all complex texture parameters viz., $a, b, c, d$ and $t$ can be expressed in terms three real texture parameters viz., $Re[d], Im[d]$ and $Re[t]$. Which means, these three independent real parameters can span $M_{\nu}$, and all the predictions associated with it.
		
		 In the upcoming section, we shall discuss the numerical predictions of the model.
\section{Numerical Analysis}
\label{section 3}
\begin{figure*}
  \centering
  \subfigure[]{\includegraphics[width=0.33
  \textwidth]{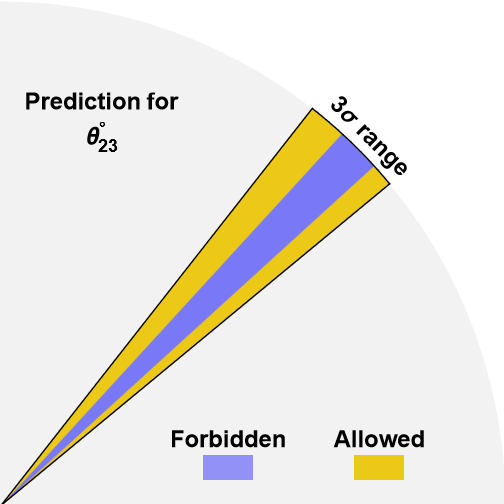}}
\caption{Constraints on $\theta_{23}$ set by our model in case of normal hierarchy.}
\label{fig:1}
\end{figure*}

We first conduct the analysis in the light of normal hierarchy of neutrino masses. The aim is to diagonalise $M_{\nu}$ with the matrix $\tilde{U}$, for which, we consider the entire $3 \sigma$ range\,\cite{Esteban:2020cvm, Gonzalez-Garcia:2021dve} of the mixing angles and $\delta$ as inputs. It is important to mention that, though we have given the $3 \sigma$ range of the mixing angles as inputs, but the model restricts $\theta_{23}$ to take some specific values within the $3\sigma$ bound. We find that there is a forbidden region approximately $42.33^{\circ}$<$\theta_{23}$<$47.50^{\circ}$ within the $3\sigma$ bound of $\theta_{23}$, which is graphically represented in Fig.\ref{fig:1}. This is one of the highlighting features of this work.


 In addition to the prediction of Majorana phases, the model predicts the three neutrino mass eigenvalues which are consistent with the experimental observables $\Delta m^2_{21}$\,\cite{Esteban:2020cvm, Gonzalez-Garcia:2021dve}, $\Delta m^2_{31}$\,\cite{Esteban:2020cvm, Gonzalez-Garcia:2021dve} and $\sum m_i< 0.12$eV~\cite{Planck:2018vyg}. The numerical values of these predictions are shown in Table\,\ref{table:2}. Furthermore, from the analysis, we also fetch the the numerical information of the free parameters, $Re[d], Im[d]$ and $Re[t]$. The free parameters $Re[d], Im[d]$ and $Re[t]$ are found to be in the ranges as shown in Table.\ref{table:3}. 

\begin{figure*}
  \centering
    \subfigure[]{\includegraphics[width=0.4952\textwidth]{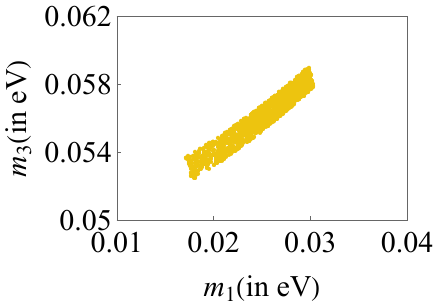}\label{fig:2(a)}}
     \subfigure[]{\includegraphics[width=0.496
  \textwidth]{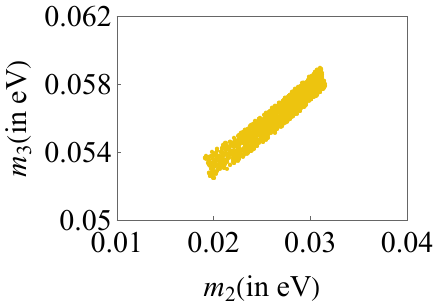}\label{fig:2(b)}}
  \subfigure[]{\includegraphics[width=0.485
  \textwidth]{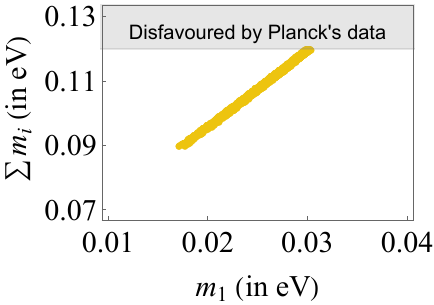}\label{fig:2(c)}}
  \subfigure[]{\includegraphics[width=0.493
  \textwidth]{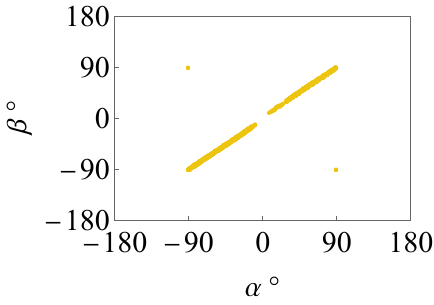}\label{fig:2(d)}}
     \subfigure[]{\includegraphics[width=0.495
  \textwidth]{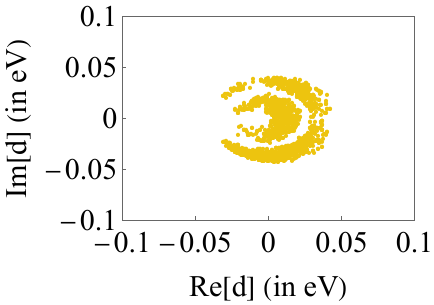}\label{fig:2(e)}} 
    \subfigure[]{\includegraphics[width=0.495\textwidth]{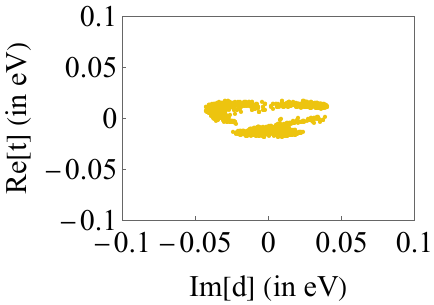}\label{fig:2(f)}} 
\caption{Represents the correlation plots for the mass eigenvalues, sum of the neutrino mass eigenvalues, Majorana phases and the free parameters in case normal hierarchy of neutrino masses.}
\label{fig:2}
\end{figure*}

To understand the predictions graphically, we show the plots in Figs. \ref{fig:2(a)}, \ref{fig:2(b)}, \ref{fig:2(c)} and \ref{fig:2(d)}. To visualize the free parameters, we highlight the Figs.\ref{fig:2(e)} and \ref{fig:2(f)}.
\begin{table}[h]
\centering
\begin{tabular}{lcc}
\hline
Prediction & Minimum & Maximum\\
\hline
$m_1 /\text{eV}$ & 0.01719 & 0.0303\\
\hline
$m_2 /\text{eV}$ & 0.01912 & 0.03155\\
\hline
$m_3 /\text{eV}$ & 0.05249& 0.05896\\
\hline
$\sum{m_i} /\text{eV}$ &0.09 &0.119\\
\hline
$\alpha$ & $-89.93^{\circ}$ & $89.77^{\circ}$\\
\hline
$\beta$ & $-89.95^{\circ}$ & $89.97^{\circ}$ \\
\hline
\end{tabular} 
\caption{Shows the numerical values of the predicted parameters in case of normal hierarchy.}
\label{table:2}
\end{table}
\begin{table}[h]
\centering
\begin{tabular}{lcc}
\hline
Free parameters & Minimum & Maximum\\
\hline
$Re[d] /\text{eV}$ & -0.030 & 0.042\\
\hline
$Im[d] /\text{eV}$ & -0.042 & 0.040\\
\hline
$Re[t] /\text{eV}$ & -0.017& 0.018\\
\hline
\end{tabular} 
\caption{Highlights the maximum and minimum values of the free parameters in case of normal hierarchy.}
\label{table:3}
\end{table}


\begin{figure*}
  \centering
 
    \subfigure[]{\includegraphics[width=0.4952\textwidth]{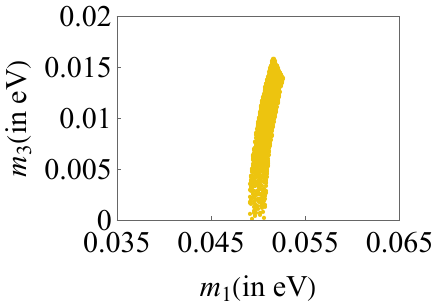}\label{fig:3(a)}}
     \subfigure[]{\includegraphics[width=0.496
  \textwidth]{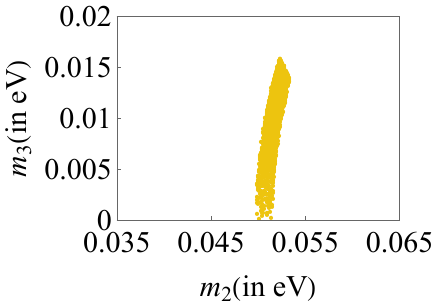}\label{fig:3(b)}}
  \subfigure[]{\includegraphics[width=0.485
  \textwidth]{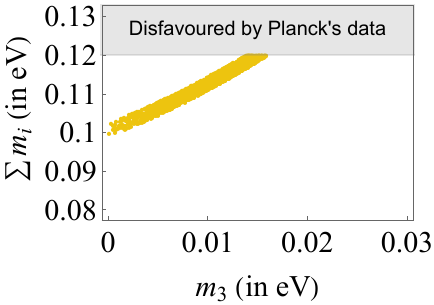}\label{fig:3(c)}}
  \subfigure[]{\includegraphics[width=0.49
  \textwidth]{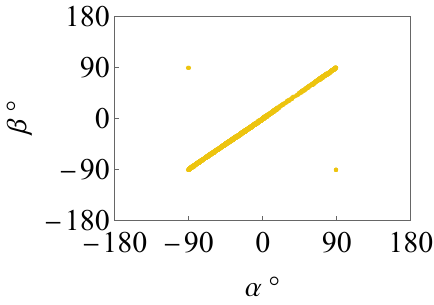}\label{fig:3(d)}}
  \subfigure[]{\includegraphics[width=0.495
  \textwidth]{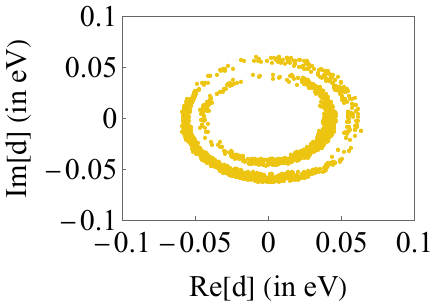}\label{fig:3(e)}} 
    \subfigure[]{\includegraphics[width=0.495\textwidth]{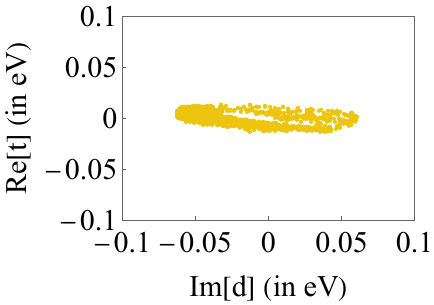}\label{fig:3(f)}}
\caption{The correlation plots for the mass eigenvalues, sum of the three neutrino mass eigenvalues, Majorana phases and the free parameters in case of inverted hierarchy of neutrino masses.}
\label{fig:3}
\end{figure*}

	For the inverted hierarchy of neutrino masses, we do the similar analysis. However, for this case, there is no constrained region for $\theta_{23}$ or the  other two mixing angles. The numerical values of three neutrino mass eigenvalues and the two Majorana phases are presented in Table \ref{table:4}. Similar to the normal hierarchy, we observe that mass eigenvalues are consistent with the observables $\Delta m^2_{21}$\,\cite{Esteban:2020cvm, Gonzalez-Garcia:2021dve}, $\Delta m^2_{31}$\,\cite{Esteban:2020cvm, Gonzalez-Garcia:2021dve} and $\sum m_i< 0.12$eV~\cite{Planck:2018vyg}. We graphically represent the predicted parameters in Figs. \ref{fig:3(a)}, \ref{fig:3(b)}, \ref{fig:3(c)} and \ref{fig:3(d)}.
	
	The free parameters $Re[d], Im[d]$ and $Re[t]$ in this case are found to be in the numerical ranges as shown in Table.\ref{table:5}. We highlight the plots for free parameters in Figs. \ref{fig:3(e)} and \ref{fig:3(f)}.

\begin{table}[h]
\centering
\begin{tabular}{lcc}
\hline
Prediction & Minimum & Maximum\\
\hline
$m_1 /\text{eV}$ & 0.0491 & 0.0526\\
\hline
$m_2 /\text{eV}$ & 0.0498 & 0.0533\\
\hline
$m_3 /\text{eV}$ & 0.00014& 0.0159\\
\hline
$\sum{m_i} /\text{eV}$ &0.099 &0.119\\
\hline
$\alpha$ & $-89.95^{\circ}$ & $89.90^{\circ}$\\
\hline
$\beta$ & $-89.87^{\circ}$ & $89.88^{\circ}$ \\
\hline
\end{tabular} 
\caption{Shows the numerical values of the predicted parameters in case of inverted hierarchy.}
\label{table:4}
\end{table}
\begin{table}[h]
\centering
\begin{tabular}{lcc}
\hline
Free parameters & Minimum & Maximum\\
\hline
$Re[d] /\text{eV}$ & -0.0589 & 0.0634\\
\hline
$Im[d] /\text{eV}$ & -0.06216 & 0.06071\\
\hline
$Re[t] /\text{eV}$ & -0.0126& 0.0140\\
\hline
\end{tabular} 
\caption{Highlights the maximum and minimum values of the free parameters in case of inverted hierarchy.}
\label{table:5}
\end{table}
\subsection*{Information of the Model Parameters}
\label{MP}
\begin{figure*}
  \centering
 
    \subfigure[]{\includegraphics[width=0.37\textwidth]{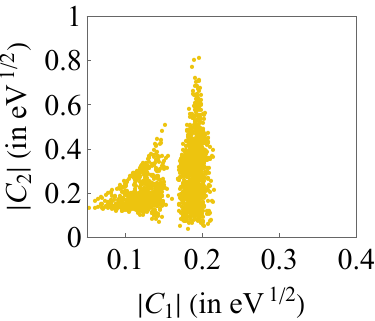}\label{fig:Cnh1}}
     \subfigure[]{\includegraphics[width=0.37
  \textwidth]{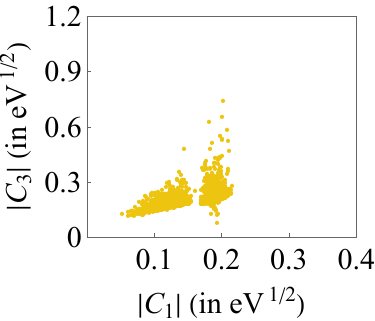}\label{fig:Cnh2}}
  \subfigure[]{\includegraphics[width=0.37
  \textwidth]{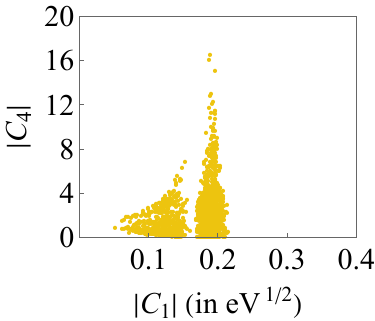}\label{fig:Cnh3}}
  \subfigure[]{\includegraphics[width=0.38
  \textwidth]{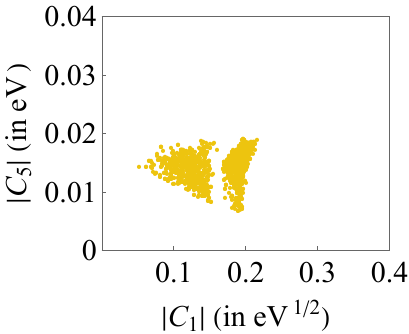}\label{fig:Cnh4}}
  \subfigure[]{\includegraphics[width=0.41
  \textwidth]{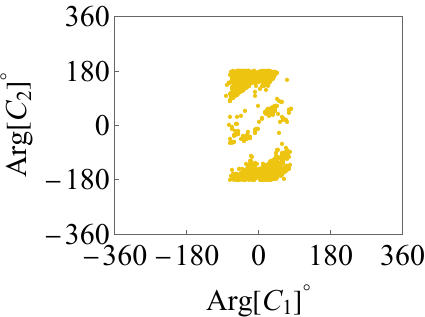}\label{fig:Cnh5}} 
    \subfigure[]{\includegraphics[width=0.41\textwidth]{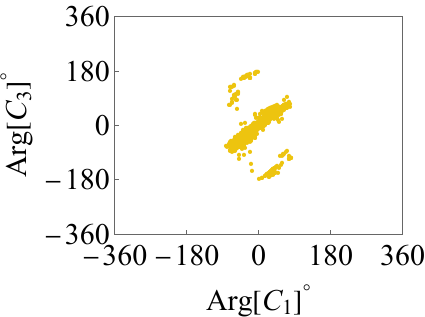}\label{fig:Cnh6}}
    \subfigure[]{\includegraphics[width=0.41
  \textwidth]{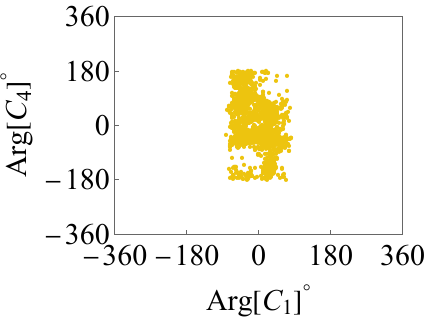}\label{fig:Cnh7}} 
    \subfigure[]{\includegraphics[width=0.41\textwidth]{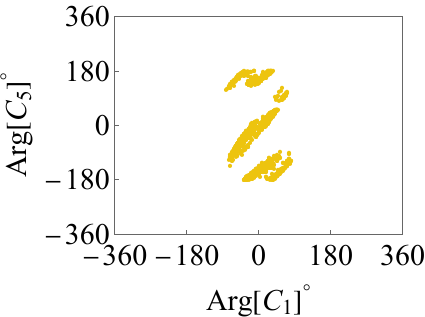}\label{fig:Cnh8}}
\caption{The correlation plots for the model parameter combinations in case of normal hierarchy of neutrino masses.}
\label{fig:Cnh}
\end{figure*}

\begin{figure*}
  \centering
    \subfigure[]{\includegraphics[width=0.35\textwidth]{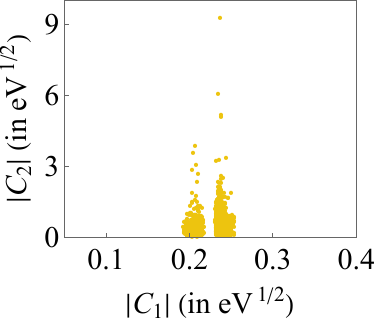}\label{fig:Cih1}}
     \subfigure[]{\includegraphics[width=0.37
  \textwidth]{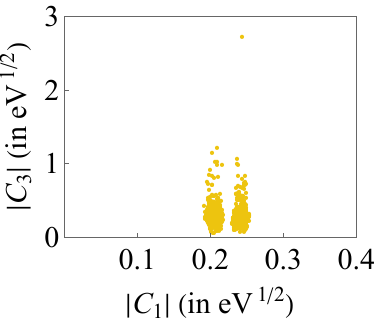}\label{fig:Cih2}}
  \subfigure[]{\includegraphics[width=0.39
  \textwidth]{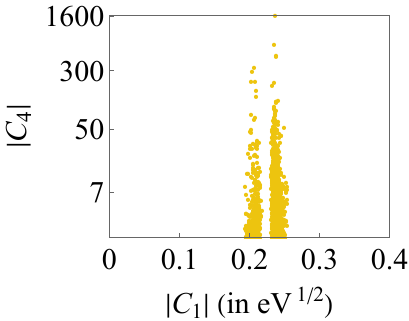}\label{fig:Cih3}}
  \subfigure[]{\includegraphics[width=0.38
  \textwidth]{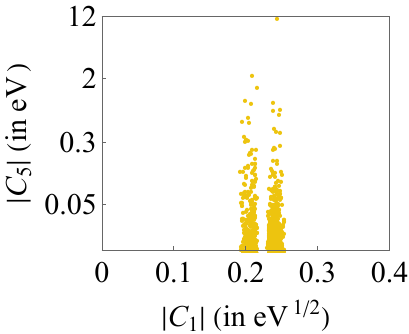}\label{fig:Cih4}}
  \subfigure[]{\includegraphics[width=0.41
  \textwidth]{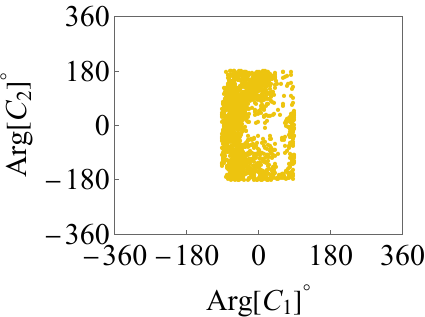}\label{fig:Cih5}} 
    \subfigure[]{\includegraphics[width=0.41\textwidth]{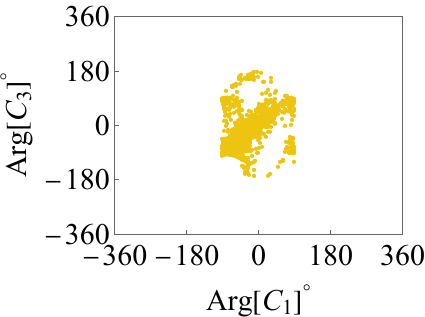}\label{fig:Cih6}}
    \subfigure[]{\includegraphics[width=0.41
  \textwidth]{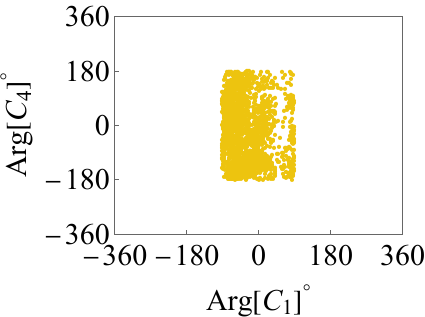}\label{fig:Cih7}} 
    \subfigure[]{\includegraphics[width=0.41\textwidth]{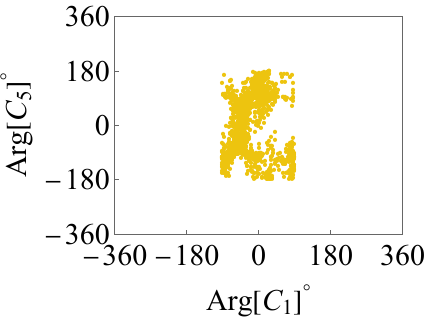}\label{fig:Cih8}}
\caption{The correlation plots for the model parameter combinations in case of inverted hierarchy of neutrino masses.}
\label{fig:Cih}
\end{figure*}

It is evident from Eq.(\ref{Model Parameters1})- Eq.(\ref{Model Parameters2}), that the model parameters can be expressed in terms of the texture parameters. However, the present model has sixteen model parameters, and we can only extract information about certain combinations of these parameters.
In this regard, we highlight the combinations: $C_1, C_2, C_3, C_4$ and $C_5$ as shown below,
\begin{eqnarray}
C_1=\frac{v_h v_{\chi }y_e }{\Lambda  \sqrt{M_1}},\quad
 C_2&=&\frac{v_h v_{\chi } v_{\xi } y_{\mu }\sqrt{y_{r_2}v_{\kappa}}}{\Lambda ^2 y_s \varrho},\quad
 C_3=\frac{v_h v_{\zeta } v_{\chi } y_{\tau }}{\Lambda ^2 \sqrt{y_{r_2}v_{\kappa}}},\nonumber\\
 C_4&=&\frac{{y_{r_1} v_{\eta} y_{r_2} v_{\kappa}}}{y^2_s \varrho^2},\quad
 C_5=v_{\Delta } y_{\Delta }.
\end{eqnarray}
It is needless to mention that these combinations are the functions of the free parameters $Re[d]$, $Im[d]$ and $Re[t]$.
Therefore, we can extract numerical information about these model parameter combinations for both hierarchies of neutrino masses. The numerical values of these combinations are crucial for calculating the branching ratio of $\mu \rightarrow e\gamma$ decay, with $C_5$ being particularly significant for this calculation. We show the numerical values of $C_1, C_2, C_3, C_4$ and $C_5$  in Table\,\ref{table:mpnh} and Table\,\ref{table:mpih}. For a graphical representation of these combinations, the plots are highlighted in Figs.\ref{fig:Cnh1}-\ref{fig:Cnh8} for the normal hierarchy and Figs.\ref{fig:Cih1}-\ref{fig:Cih8} for the inverted hierarchy.

\begin{table}[h]
\centering
\begin{tabular}{ccc}
\hline
Parameter & Minimum & Maximum\\
\hline
$|C_1|/\,\text{eV}^{\frac{1}{2}}$ & 0.052 & 0.216\\
\hline
$|C_2|/\,\text{eV}^{\frac{1}{2}}$ & 0.036 & 0.812 \\
\hline
$|C_3|/\,\text{eV}^{\frac{1}{2}}$ & 0.080 & 0.742\\
\hline
$|C_4|$ & 0.011 & 16.530\\
\hline
$|C_5|/ \text{eV} $ & 0.006 & 0.0191\\ 
\hline
$Arg[C_1]/^{\circ}$ &-79.921 &82.001\\
\hline
$Arg[C_2]/^{\circ}$ &-179.974 & 179.970\\
\hline
$Arg[C_3]/^{\circ}$ &-175.861 & 178.480\\
\hline
$Arg[C_4]/^{\circ}$ &-177.820 & 178.874\\
\hline
$Arg[C_5]/^{\circ}$ &-179.944 & 179.863\\
\hline
\end{tabular} 
\caption{ Represents the maximum and minimum values of the model parameters for normal hierarchy of neutrino masses.}
\label{table:mpnh}
\end{table}

\begin{table}[h]
\centering
\begin{tabular}{ccc}
\hline
Parameter & Minimum & Maximum\\
\hline
$|C_1|/\,\text{eV}^{\frac{1}{2}}$ & 0.192 & 0.254\\
\hline
$|C_2|/\,\text{eV}^{\frac{1}{2}}$ & 0.042 & 9.303 \\
\hline
$|C_3|/\,\text{eV}^{\frac{1}{2}}$ & 0.064 & 2.720\\
\hline
$|C_4|$ & 0.0018 & $1.64 \times 10^3$ \\
\hline
$|C_5|/ \text{eV} $ & 0.0004 & 11.144\\ 
\hline
$Arg[C_1]/^{\circ}$ &-89.996 &89.699\\
\hline
$Arg[C_2]/^{\circ}$ &-179.849 & 179.814\\
\hline
$Arg[C_3]/^{\circ}$ &-168.629 & 176.350\\
\hline
$Arg[C_4]/^{\circ}$ &-179.593 & 179.793\\
\hline
$Arg[C_5]/^{\circ}$ &-178.116 & 179.754\\
\hline
\end{tabular} 
\caption{ Represents the maximum and minimum values of the model parameters for inverted hierarchy of neutrino masses.}
\label{table:mpih}
\end{table}

The model is now equipped with the all the numerical informations necessary to study the $0 \nu \beta \beta$ decay and cLFV.
\section{Neutrinoless Double Beta Decay}
\label{section 4}
The Model adheres to the Majorana nature of the neutrinos, therefore, it is pertinent to discuss the effective Majorana neutrino mass, $m_{\beta\beta}$. The $m_{\beta \beta}$ is an important observable parameter involved in the half life\,($T_{1/2}$) of $0 \nu \beta \beta$ decay \cite{Schechter:1981bd, Blennow:2010th, Vergados:2012xy, Bilenky:2012qi, Deppisch:2012nb, Bilenky:2014uka, Dolinski:2019nrj, Agostini:2022zub, Barabash:2023dwc}. The expression for $m_{\beta\beta}$, is given by,
\begin{equation}
m_{\beta\beta} = |U^2_{ei} m_i|, \quad \text{where,}\,\, i=1,2,3.
\end{equation}	
As mentioned earlier, $ Re[d],Im[d]$ and $Re[t]$ span $M_{\nu}$ and all the predictions associated with it. Hence, $m_{\beta \beta}$ is no exception. We calculate the numerical values of $m_{\beta\beta}$ for both normal and inverted hierarchy of neutrino masses. So far, we are aware of the upper bounds concerning this parameter from various experiments\,(see Table \ref{table:6}). 
The model predicts $m_{\beta \beta}$ values that lie within the limits established by combined analysis of GERDA\,\cite{GERDA:2020xhi} and KamLAND-Zen\,\cite{KamLAND-Zen:2022tow,KamLAND-Zen:2016pfg} experiments. For normal hierarchy, our model predicts $m_{\beta \beta}$ values in the range of $0.0165$ eV to $0.0311$ eV, which is within the future sensitivity region of the LEGEND-1000 experiment \cite{LEGEND:2021bnm}. In case of inverted hierarchy, the predicted range is $0.0482$ eV to $0.0520$ eV.

\begin{figure*}
  \centering
    \subfigure[]{\includegraphics[width=0.495
  \textwidth]{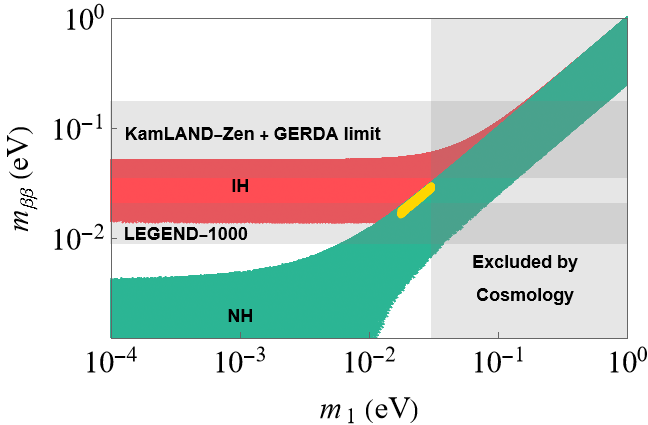}\label{fig:4(a)}} 
    \subfigure[]{\includegraphics[width=0.495\textwidth]{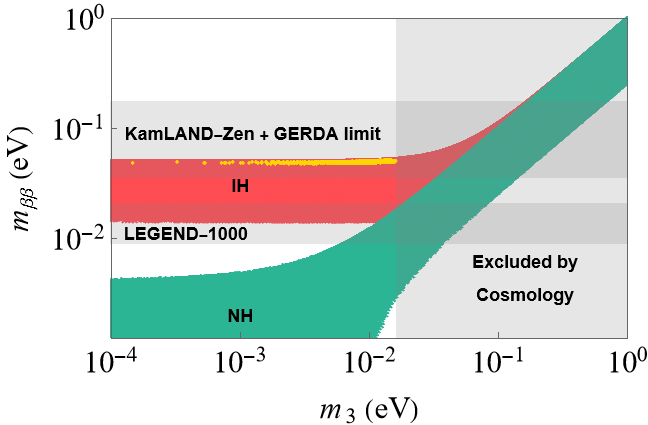}\label{fig:4(b)}} 
    \caption{The plots for $m_{\beta \beta}$ versus the lightest neutrino mass eigenvalue: (a) for normal hierarchy (b) for inverted hierarchy. In the plots, the red and green bands show the regions allowed by current oscillation data, the gray bands indicate the bounds on $m_{\beta\beta}$ and the lowest neutrino mass eigenvalues, and the `bright yellow' plot represent the model's prediction.}
\label{fig:4}
\end{figure*}

In Figs.\,\ref{fig:4(a)} and \ref{fig:4(b)}, we show the values of $m_{\beta\beta}$ versus the lightest neutrino mass eigenvalues.
\begin{table}[h]
\centering
\begin{tabular}{lccc}
\hline
Isotope & $T_{1/2}$ (years) & $m_{\beta\beta}$ (eV)& Collaboration\\
\hline
${}^{82}\mathrm{Se}$ & > $3.5 \times 10^{24}$ & $< 0.311-0.638$& CUPID-0\,\cite{CUPID:2019gpc}\\
\hline
${}^{130}\mathrm{Te}$ & > $2.2 \times 10^{25}$ & $< 0.09-0.305$& CUORE\,\cite{CUORE:2021mvw}\\
\hline
${}^{136}\mathrm{Xe}$ & > $3.5 \times 10^{25}$ & $< 0.093-0.286$& EXO\,\cite{EXO-200:2019rkq}\\
\hline
${}^{76}\mathrm{Ge}$ & > $1.8 \times 10^{26}$ & $< 0.08-0.18$& GERDA\, \cite{GERDA:2020xhi}\\
\hline
${}^{136}\mathrm{Xe}$ & > $1.07 \times 10^{26}$ & $< 0.061-0.165$& KamLAND-Zen\,\cite{KamLAND-Zen:2016pfg}\\
 & > $2.3 \times 10^{26}$ & $< 0.036-0.156$& \cite{KamLAND-Zen:2022tow}\\
\hline
${}^{76}\mathrm{Ge}$ & > $1.3 \times 10^{28}$ & $< 0.09-0.21$& LEGEND-1000\, \cite{LEGEND:2021bnm}\\
\hline
\end{tabular} 
\caption{The present lower limits on the half life\,($T_{1/2}$) and upper limits on the $m_{\beta \beta}$ of $0\nu\beta\beta$ decay for different isotopes.}
\label{table:6}
\end{table}

\section{Charged Lepton Flavour Violation}
\label{section 5}
In the SM of Particle Physics, cLFV processes\,\cite{Calibbi:2017uvl,Ardu:2022sbt,Cei:2014jtm} are incredibly rare, with predicted branching ratios around $10^{-50}$. Thus, cLFV is one of the most interesting probes of physics beyond the SM.
Among the various decay channels, transitions involving muons, are the promising ones because muons are abundantly produced in cosmic radiation and accelerators, and they have a longer lifetime than other leptons, making them ideal for study. 
In the present work, we study the dominant decay, $\mu \rightarrow e\gamma$. The current best limit on the branching ratio of this decay is set by MEG experiment\,\cite{MEG:2016leq}: $BR(\mu \rightarrow e\gamma)< 4.2 \times 10^{-13}$(90\% CL). The MEG II\,\cite{Meucci:2022qbh,MEGII:2023ltw,Francesconi:2024bpo} collaboration is searching for this decay at the Paul Scherrer Institut\,(PSI) muon beam facility.  Their goal is to improve the sensitivity of their measurements to achieve a branching ratio : $BR(\mu \rightarrow e\gamma)< 6 \times 10^{-14}$(90\% CL) by the end of 2026. To complement these experimental efforts, it's worthwhile to explore theoretical beyond standard models that could explain or predict these rare decays.

 		Having said that, as we consider the Type-I and Type-II seesaw mechanisms in our model, we may therefore analyze how each mechanism contributes to the  $\mu \rightarrow e\gamma$ decay. We then try to see which contribution is the dominant one.

In the case of Type-I seesaw, the possibility of $\mu \rightarrow e\gamma$ decay arises due to the existence of light-heavy neutrino mixing. The expression for branching ratio for the Type-I seesaw is written as\,\cite{Bilenky:1977du, Ilakovac:1994kj, Tommasini:1995ii, Forero:2011pc, Datta:2021zzf},
\begin{equation}
BR(\mu \rightarrow e\gamma)= \frac{3 \alpha_f}{8 \pi}\left| \sum_{i} K_{ei}K^{\dagger}_{i\mu} F\left(\frac{M_i^2}{M_W^2}\right) \right|^2,
\end{equation}
where, $\alpha_f$ is the fine structure constant, $K$ represents the light-heavy mixing matrix, $M_i$ is the mass of the right handed neutrinos and $M_W$ is the mass of $W^{\pm}$. The form factor $F(x)$ is expressed as,
\begin{equation}
F(x)=\frac{x(1-6x+3x^2+2x^3-6x^2 lnx)}{2(1-x)^4},
\end{equation}
where, $x=\frac{M_i^2}{M_W^2}$.
	It is important to mention that the branching ratio is dependent on $K= M_D^{\dagger}(M_R^{\dagger})^{-1} U_R$, where $U_R$ is the matrix which diagonalises $M_R$. Therefore, the contribution to this decay is dependent on the flavour structure of the neutrino mass matrices, $M_D$ and $M_R$. In our model, $M_D$ is a diagonal matrix. For there to be a significant and non vanishing contribution to the branching ratio, $M_R$ should have non-zero entries in the $(1,1)$, $(1,2)$, $(1,3)$, $(2,2)$ and $(2,3)$ positions. However, the structure of $M_R$ in our model is such that  $(1,2)$ and $(1,3)$ are vanishing entries, which forces the matrix element $K_{ei}K^{\dagger}_{i\mu}$ to  be zero. So, we conclude that the Type-I contribution to the $\mu \rightarrow e\gamma$ decay is zero in our model.

		With this inference in mind, let's now shift our focus to the Type-II seesaw contribution to the branching ratio of $\mu \rightarrow e\gamma$ decay. In the Type-II case, this process 	occurs at the one loop level, involving the doubly and singly charged components of the triplet Higgs, $\Delta$. The branching ratio for this process is given by\,\cite{Dinh:2012bp, Ferreira:2019qpf, Barrie:2022ake},
\begin{equation}
BR(\mu \rightarrow e\gamma)= \frac{\alpha_f \left| (Y^* Y)^2_{e\mu} \right|}{192 \pi G_F^2}\left(\frac{1}{m^2_{\Delta^{+}}}+ \frac{8}{m^2_{\Delta^{++}}}\right)^2.
\label{T2BR}
\end{equation}
In the above equation, $Y$ is the Type-II Yukawa  matrix, $G_F$ is the Fermi constant, $m_{\Delta^{+}}$	is the mass of the singly charged component of the triplet field and $m_{\Delta^{++}}$ stands for the mass of the doubly charged component of the triplet field.
In our analysis we assume small differences between the components of $\Delta$, which means, $m_{\Delta^{+}} \simeq m_{\Delta^{++}}= m_{\Delta}$. With this assumption and using the value of $ (Y^* Y)_{e\mu}$
from the model, the branching ratio takes the following form,
\begin{equation}
BR(\mu \rightarrow e\gamma)\simeq 0.659 \frac{\alpha_f}{\pi G_F^2}\left(\frac{ \left|C_5\right|}{m_{\Delta} v_{\Delta}}\right)^4.
\label{T2BRM}
\end{equation}

From Eq.,(\ref{T2BRM}), we understand that the branching ratio depends on the Type-II parameter, $C_5$. Given the experimental sensitivities of $BR(\mu \rightarrow e\gamma)$ and the values of $C_5$ predicted by this model, we determine the possible numerical ranges of $m_{\Delta}$ and $v_{\Delta}$. These values are summarized in Table \ref{table:clfv}.
\begin{table}[h]
    \centering
    \begin{tabular}{c cc cc}
        \hline
        \multirow{2}{*}{Parameter} & \multicolumn{2}{c}{Normal hierarchy} & \multicolumn{2}{c}{Inverted hierarchy} \\
        \cline{2-5}
         & Minimum & Maximum  & Minimum & Maximum \\
        \hline
        $BR(\mu \rightarrow e\gamma)$ & $10^{-16}$ & $7.70 \times 10^{-13}$  & $10^{-16}$ & $6.94 \times 10^{-13}$ \\
        \hline
        $m_{\Delta}$ & 140 GeV & $10^4$ GeV  & 123 GeV & $10^4$ GeV \\
        \hline
        $v_{\Delta}$ & 1 eV & 6.97 eV  & 1 eV & 6.98 eV \\
        \hline
        $|C_5|$ & 0.0078 eV & 0.0191 eV  & 0.0026 eV & 0.7770 eV \\
        \hline
    \end{tabular}
    \caption{Table with the model predictions for the $\mu \rightarrow e\gamma$ decay in case of normal and inverted hierarchies.}
    \label{table:clfv}
    \end{table}

\begin{figure*}
  \centering
    \subfigure[]{\includegraphics[width=0.495
  \textwidth]{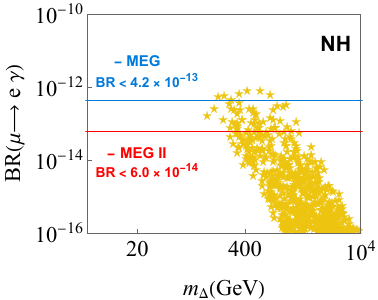}\label{fig:5(a)}}
    \subfigure[]{\includegraphics[width=0.495\textwidth]{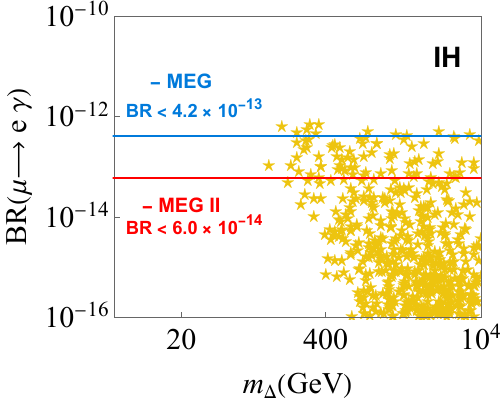}\label{fig:5(b)}} 
    \caption{$BR(\mu \rightarrow e\gamma)$ versus $m_{\Delta}$, where $m_{\Delta}= m_{\Delta^{++}} \simeq m_{\Delta^{+}}$ in this model: (a) for normal hierarchy (b) for inverted hierarchy. In the plots, blue and red lines represent the sensitivities of MEG and MEG-II respectively. The `dark yellow' stars ($\star$) are the predictions of our model.}
\label{fig:5}
\end{figure*}

Based on the values in Table \ref{table:clfv}, we understand that the $BR(\mu \rightarrow e\gamma)$ predicted by our model is consistent with the upper bounds set by the MEG and MEG-II experiments, suggesting that $m_{\Delta}$ lies approximately between 123 GeV and  $10^4$ GeV (1 TeV). This implies that detecting the decay in the ongoing searches at MEG-II \cite{Meucci:2022qbh,MEGII:2023ltw,Francesconi:2024bpo} experiment could set limits on the masses of doubly or singly charged scalars, likely between 123 GeV and  $10^4$ GeV (1 TeV). To better understand these findings, we highlight the Figs. \ref{fig:5(a)} and \ref{fig:5(b)}.	
	
\section{Summary and Discussions}
\label{section 6}
In this paper, we have constructed a neutrino mass matrix model based on the extension of the Standard Model with $\Delta(27)$ symmetry in a Type-I + Type-II framework.
 An additional $Z_{10}$ symmetry has been introduced to prevent certain unfavourable terms from appearing in the Yukawa lagrangian. The mass matrix in our model exhibits a broken $\mu$-$\tau$ symmetry and contains only \emph{three real and independent texture parameters} that collectively describe the entire mass matrix and its associated predictions.

To derive the phenomenology associated with the texture, we have conducted our analysis for both normal and inverted hierarchies of neutrino masses. In this regard, we have used the $3 \sigma$ values for the mixing angles and the Dirac CP phase as inputs. However, we have observed that the model naturally restricts $\theta_{23}$ to specific ranges of values in the case of normal hierarchy. This is one of the highlighting features of our model.

Furthermore, our model predicts the three neutrino mass eigenvalues and allows the two Majorana phases to vary approximately between $-90^{\circ}$ and $90^{\circ}$. To test our model on experimental grounds, we have extended our discussion in the light of neutrino-less double beta ($0\nu\beta\beta$) decay and charged lepton flavour violation (cLFV). Our model predicts the highest possible values for $m_{\beta\beta}$ to be $0.0311$ eV and $0.0520$ eV for normal and inverted hierarchies, respectively, which are consistent with the experimental upper bounds.

In the context of cLFV, we explore our model's contribution to the $\mu \rightarrow e \gamma$ decay. In this study, we observe an interesting result: due to the presence of texture zeros in the structure of $M_R$, the Type-I seesaw mechanism provides no contribution to this decay. Only the Type-II seesaw mechanism offers a non-zero contribution to the branching ratio, and we have explored the values of $m_{\Delta}$ for which the branching ratio is consistent with current and future experimental sensitivities. From the study we have found that $m_{\Delta}$ lies approximately between 123 GeV and  $10^4$ GeV (1 TeV).

\section*{Acknowledgements}
MD would like to thank R. Srivastava, Department of Physics, IISER-Bhopal, for an important discussion related to the $\Delta(27)$ symmetry group, and B. Karmakar, University of Silesia, Katowice, for a few discussions related to cLFV. The research work of MD is financially supported by Council of Scientific and Industrial Research\,(CSIR), Government of India through a NET Junior Research Fellowship vide grant No. 09/0059(15346)/2022-EMR-I.



\bibliographystyle{JHEP}
\bibliography{reference1.bib}
\appendix
\section{Appendix}
\label{Appendix}
\subsection{The Pontecorvo-Maki-Nakagawa-Sakata matrix}
The Pontecorvo-Maki-Nakagawa-Sakata(PMNS) matrix is a $3 \times 3$ unitary matrix which is parametrised in terms of mixing angles and phases. The parametrisation of the PMNS matrix\,($U$) adopted by the Particle Data Group ~\cite{ParticleDataGroup:2018ovx} is shown below,
\begin{equation}
\label{pmns}
U= P_{\phi}.\tilde{U}. P_M ,
\end{equation}
we demarcate $U$ which is the general one, from the $\tilde{U}$ that excludes the phases, where, $P_M=diag(e^{i\alpha},e^{i\beta},1)$ contains the two Majorana phases, $\alpha$ and $\beta$. The arbitrary phase matrix $P_{\phi} =diag (e^{i\phi_1},e^{i\phi_2},e^{i\phi_3})$, contains three arbitrary phases\,($\phi_{i=1,2,3}$), and the latter can be eliminated from $U$ by redefining the charged lepton fields in terms of these phases. Hence, the PMNS matrix after redefinition of the said phases, attains the following form,
\begin{equation}
\label{pmns1}
U= \tilde{U}. P_M .
\end{equation} 
The matrix $\tilde{U}$ is written in its standard form, in the following manner,
\begin{eqnarray}
\tilde{U} &=& \begin{bmatrix}
1 & 0 & 0\\
0 & c_{23} & s_{23}\\
0 & - s_{23} & c_{23}
\end{bmatrix}\times \begin{bmatrix}
c_{13} & 0 & s_{13}\,e^{-i\delta}\\
0 & 1 & 0\\
-s_{13} e^{i\delta} & 0 & c_{13}
\end{bmatrix}\times\begin{bmatrix}
c_{12} & s_{12} & 0\\
-s_{12} & c_{12} & 0\\
0 & 0 & 1
\end{bmatrix},
\end{eqnarray}
where, $s_{ij}$ and $c_{ij}$ represents $\sin\theta_{ij}$ and $\cos\theta_{ij}$ respectively. 
If the charged lepton mass matrix is already diagonal, then usually, the matrix $U$ is the diagonalising matrix for $M_{\nu}$, i.e,
\begin{equation}
U^T. M_{\nu}.\,U= diag(m_1, m_2, m_3),
\label{11}
\end{equation}
However, we rearrange the above equation in the following way,
\begin{equation}
\tilde{U}^T. M_{\nu}.\,\tilde{U}= diag(\tilde{m_1},\tilde{m_2},m_3),
\label{13}
\end{equation}
such that, $\tilde{m_1} = m_1 e^{-2 i \alpha}$ and $\tilde{m_2} = m_2 e^{-2 i \beta}$. In the present model, $\tilde{U}$ is the PMNS matrix which diagonalises the neutrino mass matrix $M_{\nu}$.

\subsection{$\Delta(27)$ group}
We briefly explore the product rules associated with $\Delta(27)$ symmetry,
\begin{eqnarray}
3 \otimes 3 &=& \bar{3}_{s_1} \oplus \bar{3}_{s_2} \oplus  \bar{3}_a, \nonumber\\
\bar{3} \otimes \bar{3} &=& 3_{s_1} \oplus  3_{s_2} \oplus  3_a, \nonumber\\
3 \otimes \bar{3} &=& \sum_{r=0}^2 1_{r,0} \oplus  \sum_{r=0}^2 1_{r,1} \oplus  \sum_{r=0}^2 1_{r,2},\nonumber\\
1_{p,q} \otimes 1_{p',q'} &=& 1_{(p+p')\,\text{mod}\,3,\, (q+q')\,\text{mod}\,3}.
\end{eqnarray}
If $(a_1, a_2, a_3)$ and $(b_1, b_2, b_3)$ are two triplets under $\Delta(27)$ then,
\begin{eqnarray}
(3 \otimes 3)_{\bar{3}_{s_1}} &=& (a_1 b_1 , a_2 b_2 , a_3 b_3),\nonumber\\
(3 \otimes 3)_{\bar{3}_{s_2}} &=& \frac{1}{2}(a_2 b_3 + a_3 b_2, a_3 b_1 + a_1 b_3, a_1 b_2 + a_2 b_1),\nonumber\\
(3 \otimes 3)_{\bar{3}_a} &=& \frac{1}{2}(a_2 b_3 - a_3 b_2, a_3 b_1 - a_1 b_3, a_1 b_2 - a_2 b_1),\nonumber\\
(3 \otimes \bar{3})_{1_{r,0}} &=& a_1 b_1 + \omega^{2r} a_2 b_2 + \omega^r a_3 b_3,\nonumber\\
(3 \otimes \bar{3})_{1_{r,1}} &=& a_1 b_2 + \omega^{2r} a_2 b_3 + \omega^r a_3 b_1, \nonumber\\
(3 \otimes \bar{3})_{1_{r,2}} &=& a_1 b_3 + \omega^{2r} a_2 b_1 + \omega^r a_3 b_2, 
\end{eqnarray}
where, $r= 0,1,2$ and $\omega= e^{\frac{2\pi i}{3}}$.
\subsection{$Z_{10}$ group}
$Z_{10}$ group, represents a modulus symmetry and it involves the elements 0 to 9.
If, $x_1$ and $x_2$ are two group elements of $Z_{10}$, then under group operation:
\begin{equation}
x_1 \times x_2 = (x_1+x_2)\,mod\,10.
\end{equation}
The irreducible representation of a general group element\,($x$), is given by: $e^{2 x \pi i/10}$. In Table\,\ref{table:z10}, we show the multiplication table of the $Z_{10}$ group,
\begin{table}[h]
\centering
\begin{tabular}{|p{0.4cm}|p{0.1cm}p{0.1cm}p{0.1cm}p{0.1cm}p{0.1cm}p{0.1cm}p{0.1cm}p{0.1cm}p{0.1cm}p{0.1cm}|}
\hline
$Z_{10}$& \,0 & 1 & 2 & 3 & 4 & 5 & 6 & 7 & 8 & 9 \\
 \hline
0& \,0 & 1 & 2 & 3 & 4 & 5 & 6 & 7 & 8 & 9\\
\hline
1& \,1 & 2 & 3 & 4 & 5 & 6 & 7 & 8 & 9 & \textbf{0} \\
\hline
2& \,2 & 3 & 4 & 5 & 6 & 7 & 8 & 9 & \textbf{0} & 1 \\
\hline
3& \,3 & 4 & 5 & 6 & 7 & 8 & 9 & \textbf{0}& 1 & 2 \\
\hline
4& \,4 & 5 & 6 & 7 & 8 & 9 & \textbf{0} & 1 & 2 & 3 \\
\hline
5& \,5 & 6 & 7 & 8 & 9 & \textbf{0} & 1 & 2 & 3 & 4 \\
\hline
6& \,6 & 7 & 8 & 9 & \textbf{0} & 1 & 2 & 3 & 4 & 5 \\
\hline
7& \,7 & 8 & 9 & \textbf{0} & 1 & 2 & 3 & 4 & 5 & 6 \\
\hline
8& \,8 & 9 & \textbf{0} & 1 & 2 & 3 & 4 & 5 & 6 & 7 \\
\hline
9& \,9 & \textbf{0} & 1 & 2 & 3 & 4 & 5 & 6 & 7 & 8 \\
\hline
\end{tabular}
\caption{Represents the multiplication table of the $Z_{10} $ group.} 
\label{table:z10}
\end{table}
\subsection{Scalar Potential}
The scalar potential of our model which is invariant under $SU(2)_{L}\times \Delta(27) \times Z_{10}$ is presented as shown:
\begin{eqnarray}
V&=& V(H)+ V(\chi)+V(\Delta)+V(\eta)+ V(\kappa)+V(\xi)+V(\zeta)+V(\varrho)+V(interaction).\nonumber
\end{eqnarray}
Writing the terms explicitly, we have,
\begin{align}
V(H)&= -\mu^2_{H}(H^{\dagger}H)+ \lambda^{H}(H^{\dagger}H)(H^{\dagger}H),\nonumber\\
V(\chi)&=\mu^2_{\chi}(\chi^{\dagger}\chi)+\lambda^{\chi}_1(\chi^{\dagger}\chi)(\chi^{\dagger}\chi)+\lambda^{\chi}_2(\chi^{\dagger}\chi)_{1_{10}}(\chi^{\dagger}\chi)_{1_{20}}+\lambda^{\chi}_3(\chi^{\dagger}\chi)_{1_{01}}(\chi^{\dagger}\chi)_{1_{02}}+\lambda^{\chi}_4(\chi^{\dagger}\nonumber\\&\quad\chi)_{1_{11}}(\chi^{\dagger}\chi)_{1_{22}}+\lambda^{\chi}_5(\chi^{\dagger}\chi)_{1_{21}}(\chi^{\dagger}\chi)_{1_{12}},\nonumber\\
V(\Delta)&=\mu^2_{\Delta}Tr(\Delta^{\dagger}\Delta)+ \lambda^{\Delta}_1 Tr(\Delta^{\dagger}\Delta)Tr(\Delta^{\dagger}\Delta)+\lambda^{\Delta}_2 Tr(\Delta^{\dagger}\Delta)_{1_{10}} Tr(\Delta^{\dagger}\Delta)_{1_{20}}+\lambda^{\Delta}_3 Tr(\nonumber\\&\quad\Delta^{\dagger}\Delta)_{1_{01}} Tr(\Delta^{\dagger}\Delta)_{1_{02}}+\lambda^{\Delta}_4 Tr(\Delta^{\dagger}\Delta)_{1_{11}} Tr(\Delta^{\dagger}\Delta)_{1_{22}}+\lambda^{\Delta}_5 Tr(\Delta^{\dagger}\Delta)_{1_{21}} Tr(\Delta^{\dagger}\Delta)_{1_{12}},\nonumber\\
V(\eta)&=  \mu^2_{\eta}(\eta^{\dagger}\eta)+ \lambda^{\eta}(\eta^{\dagger}\eta)^2,\nonumber\\
V(\kappa)&=  \mu^2_{\kappa}(\kappa^{\dagger}\kappa)+ \lambda^{\kappa}(\kappa^{\dagger}\kappa)^2,\nonumber\\
V(\xi)&=  \mu^2_{\xi}(\xi^{\dagger}\xi)+ \lambda^{\xi}(\xi^{\dagger}\xi)^2,\nonumber\\
V(\zeta)&=  \mu^2_{\zeta}(\zeta^{\dagger}\zeta)+ \lambda^{\zeta}(\zeta^{\dagger}\zeta)^2,\nonumber\\
V(\varrho)&= \mu^2_{\varrho}(\varrho^{\dagger}\varrho)+ \lambda^{\varrho}(\varrho^{\dagger}\varrho)^2,\nonumber
\end{align}
and the interaction terms are,
\begin{align}
V(H,\chi)&=\lambda^{H \chi}(H^{\dagger}H)(\chi^{\dagger}\chi),\nonumber\\
V(H,\Delta)&=\lambda^{H \Delta}(H^{\dagger}H)\,Tr(\Delta^{\dagger}\Delta),\nonumber\\
V(H,\eta)&=\lambda^{H \eta}(H^{\dagger}H)(\eta^{\dagger}\eta),\nonumber\\
V(H,\kappa)&=\lambda^{H \chi}(H^{\dagger}H)(\kappa^{\dagger}\kappa),\nonumber\\
V(H,\xi)&=\lambda^{H \xi}(H^{\dagger}H)(\xi^{\dagger}\xi),\nonumber\\
V(H,\zeta)&=\lambda^{H \zeta}(H^{\dagger}H)(\zeta^{\dagger}\zeta),\nonumber\\
V(H,\varrho)&=\lambda^{H \varrho}(H^{\dagger}H)(\varrho^{\dagger}\varrho),\nonumber\\
V(\chi,\Delta)&= \lambda^{\chi \Delta}_1(\chi^{\dagger}\chi)(\Delta^{\dagger}\Delta)+\lambda^{\chi \Delta}_2(\chi^{\dagger}\chi)_{1_{10}}(\Delta^{\dagger}\Delta)_{1_{20}}+\lambda^{\chi \Delta}_3(\chi^{\dagger}\chi)_{1_{01}}(\Delta^{\dagger}\Delta)_{1_{02}}+\lambda^{\chi \Delta}_4(\chi^{\dagger}\nonumber\\&\quad\chi)_{1_{11}}(\Delta^{\dagger}\Delta)_{1_{22}}+\lambda^{\chi \Delta}_5(\chi^{\dagger}\chi)_{1_{21}}(\Delta^{\dagger}\Delta)_{1_{12}}+\lambda^{\chi \Delta}_6(\chi^{\dagger}\chi)_{1_{20}}(\Delta^{\dagger}\Delta)_{1_{10}}+\lambda^{\chi \Delta}_7(\chi^{\dagger}\chi)_{1_{02}}\nonumber\\&\quad(\Delta^{\dagger}\Delta)_{1_{01}}+\lambda^{\chi \Delta}_8(\chi^{\dagger}\chi)_{1_{22}}(\Delta^{\dagger}\Delta)_{1_{11}}+\lambda^{\chi \Delta}_9(\chi^{\dagger}\chi)_{1_{12}}(\Delta^{\dagger}\Delta)_{1_{21}},\nonumber\\
V(\chi,\eta)&=\lambda^{\chi\eta}(\chi^{\dagger}\chi)\eta^{\dagger}\eta,\nonumber\\
V(\chi,\kappa)&=\lambda^{\chi \kappa}(\chi^{\dagger}\chi)\kappa^{\dagger}\kappa,\nonumber\\
V(\chi,\xi)&=\lambda^{\chi \xi}(\chi^{\dagger}\chi)\xi^{\dagger}\xi,\nonumber\\
V(\chi,\zeta)&=\lambda^{\chi \zeta}(\chi^{\dagger}\chi)\zeta^{\dagger}\zeta,\nonumber\\
V(\chi,\varrho)&=\lambda^{\chi \varrho}(\chi^{\dagger}\chi)\varrho^{\dagger}\varrho,\nonumber\\
V(\Delta,\eta)&=\lambda^{\eta \Delta }Tr(\Delta^{\dagger}\Delta)\eta^{\dagger}\eta,\nonumber\\
V(\Delta,\kappa)&=\lambda^{\kappa \Delta}Tr(\Delta^{\dagger}\Delta)\kappa^{\dagger}\kappa,\nonumber\\
V(\Delta,\xi)&=\lambda^{\Delta \xi}Tr(\Delta^{\dagger}\Delta)\xi^{\dagger}\xi,\nonumber\\
V(\Delta,\zeta)&=\lambda^{\Delta \zeta}Tr(\Delta^{\dagger}\Delta)\zeta^{\dagger}\zeta,\nonumber\\
V(\Delta,\varrho)&=\lambda^{\Delta \varrho}Tr(\Delta^{\dagger}\Delta)\varrho^{\dagger}\varrho,\nonumber\\
V(\eta,\kappa)&=\lambda^{\eta \kappa}(\eta^{\dagger}\eta \kappa^{\dagger}\kappa),\nonumber\\
V(\eta,\xi)&=\lambda^{\eta\xi}\eta^{\dagger}\eta \xi^{\dagger}\xi,\nonumber\\
V(\eta,\zeta)&=\lambda^{\eta\zeta}\eta^{\dagger}\eta \zeta^{\dagger}\zeta,\nonumber\\
V(\eta,\varrho)&=\lambda^{\eta\varrho}\eta^{\dagger}\eta \varrho^{\dagger}\varrho,\nonumber\\
V(\kappa,\xi)&=\lambda^{\kappa\xi}\kappa^{\dagger}\kappa \xi^{\dagger}\xi,\nonumber\\
V(\kappa,\zeta)&=\lambda^{\kappa\zeta}\kappa^{\dagger}\kappa \zeta^{\dagger}\zeta,\nonumber\\
V(\kappa,\varrho)&=\lambda^{\kappa\varrho}\kappa^{\dagger}\kappa \varrho^{\dagger}\varrho,\nonumber\\
V(\xi,\zeta)&=\lambda^{\xi \zeta}(\xi^{\dagger}\xi \zeta^{\dagger}\zeta),\nonumber\\
V(\xi,\varrho)&=\lambda^{\xi \varrho}(\xi^{\dagger}\xi \varrho^{\dagger}\varrho),\nonumber\\
V(\zeta,\varrho)&=\lambda^{\zeta \varrho}(\zeta^{\dagger}\zeta \varrho^{\dagger}\varrho),\nonumber\\
V(H,\Delta, \chi)&=(H^T i \sigma_2 \Delta^{T} H \chi +h.c).\nonumber
\end{align}
The term $V(H,\Delta, \chi)$ is important, as it is responsible for the vev of the Higgs triplet field, $\Delta$. In models with discrete symmetry, it is common to have multiple coupling constants in the scalar potential. This allows for the flexibility to choose appropriate vacuum alignments for the scalar fields. Without loss of generality, for the chosen vevs in the directions: $\langle \chi \rangle = v_{\chi}(1,0,0)$, $\langle H \rangle = v_h, \langle \Delta \rangle = v_{\Delta}(1,1,1), \langle \eta \rangle = v_{\eta}, \langle \kappa \rangle = v_{\kappa}, \langle \xi \rangle = v_{\xi}$, $\langle \zeta \rangle = v_{\zeta}$ and $\langle \varrho \rangle = v_{\varrho}$, and using $\lambda _6^{\chi \Delta }=\lambda _2^{\chi \Delta }$, we have the following minimisation conditions of the scalar potential,
\begin{align}
\frac{\partial V}{\partial H}&=v_h(2 v_h^2 \lambda ^{H}+3 \lambda ^{H \Delta} v_{\Delta }^2+v_\zeta^2 \lambda ^{H \zeta}+v_\eta^2 \lambda ^{H \eta}+v_\kappa^2 \lambda ^{H \kappa}+v_\xi^2 \lambda ^{H \xi}+v_\chi^2 \lambda ^{H \chi}+v_\varrho^2 \lambda ^{H \varrho}-\nonumber\\&\quad\,\,\mu_H^2-4 v_\chi v_{\Delta }) =0,\nonumber\\
\frac{\partial V}{\partial v_{\chi_1}}&=v_h^2 \left(v_{\chi} \lambda ^{H \chi}-2 v_{\Delta }\right)+v_{\chi}(\mu _{\chi }^2+3 \lambda_1 ^{\chi \Delta } v_{\Delta }^2+2 v_{\chi}^2 \lambda _1^{\chi }+2 v_{\chi}^2 \lambda _2^{\chi }+v_{\eta}^2 \lambda ^{\chi \eta }+v_{\kappa}^2 \lambda^{\chi \kappa }+v_{\xi}^2 \nonumber\\&\quad\,\,\lambda ^{\chi \xi }+v_{\zeta}^2 \lambda ^{\chi \zeta }+v_\varrho^2 \lambda^{\chi \varrho})=0,\nonumber\\
\frac{\partial V}{\partial v_{\chi_2}}&=3 v_{\chi} \lambda _3^{\chi \Delta } v_{\Delta }^2=0,\nonumber\\
\frac{\partial V}{\partial v_{\chi_3}}&=3 v_{\chi} \lambda _7^{\chi \Delta } v_{\Delta }^2=0,\nonumber\\
\frac{\partial V}{\partial v_{\Delta_1}}&=v_h^2(\lambda ^{H \Delta} v_{\Delta }-2 v_{\chi})+v_{\Delta }(\mu _{\Delta}^2+6(\lambda _1^{\Delta }+\lambda _3^{\Delta }) v_{\Delta }^2+v_{\chi}^2 \lambda _1^{\chi \Delta }+2 v_{\chi}^2 \lambda _2^{\chi \Delta }+v_{\eta}^2 \lambda^{\Delta \eta }+v_{\kappa}^2\nonumber\\&\quad\,\, \lambda^{\Delta \kappa}+v_{\xi}^2 \lambda^{\Delta \xi }+v_{\zeta}^2 \lambda^{\Delta \zeta }+v_{\varrho}^2 \lambda^{\Delta \varrho})=0,\nonumber\\
\frac{\partial V}{\partial v_{\Delta_2}}&=\frac{\partial V}{\partial v_{\Delta_3}}=v_{\Delta }(\mu _{\Delta }^2+v_h^2 \lambda^{H\Delta}+6(\lambda _1^{\Delta }+\lambda _3^{\Delta }) v_{\Delta }^2+v_{\chi}^2 \lambda _1^{\chi \Delta }-v_{\chi}^2 \lambda _2^{\chi \Delta }+v_{\eta}^2 \lambda^{\Delta \eta }+v_{\kappa}^2 \lambda ^{\Delta \kappa }\nonumber\\&\quad\,\,+v_{\xi}^2 \lambda ^{\Delta \xi }+v_{\zeta}^2 \lambda^{\Delta \zeta }+v_{\varrho}^2 \lambda^{\Delta \varrho })=0,\nonumber\\
\frac{\partial V}{\partial v_{\eta}}&=v_\eta(v_h^2 \lambda ^{H \eta }+\mu_\eta^2+3 \lambda ^{\Delta \eta } v_{\Delta }^2+v_\chi^2 \lambda ^{\chi \eta }+2 v_\eta^2 \lambda ^{\eta}+v_\kappa^2 \lambda ^{\eta \kappa }+v_\varrho^2 \lambda ^{\eta \varrho }+v_\xi^2 \lambda ^{\eta \xi }+v_\zeta^2 \lambda ^{\eta \zeta })\nonumber\\&=0,\nonumber\\
\frac{\partial V}{\partial v_{\kappa}}&=v_\kappa(v_h^2 \lambda ^{H \kappa}+\mu_\kappa ^2+3 \lambda ^{\Delta \kappa } v_{\Delta }^2+v_\chi^2 \lambda ^{\chi \kappa }+v_\eta^2 \lambda ^{\eta \kappa }+2 v_\kappa^2 \lambda ^{\kappa}+v_\varrho^2 \lambda ^{\kappa \varrho }+v_\xi^2 \lambda ^{\kappa \xi }+v_\zeta^2 \lambda ^{\kappa \zeta })\nonumber\\&=0,\nonumber\\
\frac{\partial V}{\partial v_{\xi}}&=v_\xi(v_h^2 \lambda ^{H \xi}+3 \lambda ^{\Delta \xi } v_{\Delta }^2+v_\chi^2 \lambda ^{\chi \xi }+v_\eta^2 \lambda ^{\eta \xi }+v_\kappa^2 \lambda ^{\kappa \xi }+v_\varrho^2 \lambda ^{\xi \varrho }+2 v_\xi^2 \lambda ^{\xi}+v_\zeta^2 \lambda ^{\xi \zeta }+\mu_\xi ^2)\nonumber\\&=0,\nonumber\\
\frac{\partial V}{\partial v_{\zeta}}&=v_\zeta(v_h^2 \lambda ^{H \zeta}+\mu_\zeta^2+3 \lambda ^{\Delta \zeta } v_{\Delta }^2+v_\chi^2 \lambda ^{\chi \zeta }+v_\eta^2 \lambda ^{\eta \zeta }+v_\kappa^2 \lambda ^{\kappa \zeta }+v_\varrho^2 \lambda ^{\zeta \varrho }+v_\xi^2 \lambda ^{\xi \zeta }+2 v_\zeta^2 \lambda ^{\zeta})\nonumber\\&=0,\nonumber\\
\frac{\partial V}{\partial v_{\varrho}}&=v_{\varrho}(v_h^2 \lambda ^{H \varrho}+\mu_\varrho^2+3 \lambda ^{\Delta \varrho } v_{\Delta }^2+v_\chi^2 \lambda ^{\chi \varrho }+v_\eta^2 \lambda ^{\eta \varrho }+v_\kappa^2 \lambda ^{\kappa \varrho }+2 v_{\varrho}^2 \lambda ^{\varrho}+v_\xi^2 \lambda ^{\xi \varrho }+v_\zeta^2 \lambda ^{\zeta \varrho })\nonumber\\&=0.\nonumber
\end{align}
Using the above minimisation equations, we further obtain:
\begin{align}
\mu_H&= (2 v_h^2 \lambda ^{H}+3 \lambda ^{H \Delta} v_{\Delta }^2+v_\zeta^2 \lambda ^{H \zeta}+v_\eta^2 \lambda ^{H \eta}+v_\kappa^2 \lambda ^{H \kappa}+v_\xi^2 \lambda ^{H \xi}+v_\chi^2 \lambda ^{H \chi}+v_\varrho^2 \lambda ^{H \varrho}-4 v_\chi\nonumber\\&\quad\,\, v_{\Delta })^{1/2},\\
\mu _{\chi }&=((v_h^2(2 v_{\Delta }-v_\chi \lambda ^{H \chi})-v_\chi (3 \lambda _1^{\chi \Delta } v_{\Delta }^2+2 v_\chi^2 \lambda _1^{\chi }+2 v_\chi^2 \lambda _2^{\chi }+v_\eta^2 \lambda ^{\chi \eta }+v_\kappa^2 \lambda ^{\chi \kappa }+v_\varrho^2\lambda ^{\chi \varrho }+\nonumber\\&\quad\,\,v_\xi^2  \lambda ^{\chi \xi }+v_\zeta^2 \lambda ^{\chi \zeta }))/v_{\chi})^{1/2},
\end{align}
\begin{align}
\mu _{\Delta }&= (-v_\chi^2 v_h^2 \lambda ^{H \Delta}-\frac{8}{3} (\lambda _1^{\Delta }+\lambda _3^{\Delta }) v_h^4 \lambda _2^{-2 \chi \Delta }+v_\chi^4 \lambda _2^{\chi \Delta }-v_\chi^2(v_\chi^2 \lambda _1^{\chi \Delta }+v_\eta^2 \lambda ^{\Delta \eta }+v_\kappa^2 \lambda ^{\Delta \kappa }+\nonumber\\&\quad\,\,v_\varrho^2 \lambda ^{\Delta \varrho }+v_\xi^2 \lambda ^{\Delta \xi }+v_\zeta^2 \lambda ^{\Delta \zeta }))^{1/2}/v_{\chi},\\
\label{33}
v_{\Delta }&= \frac{2 v_h^2}{3 \lambda _2^{\chi \Delta }v_{\chi}},\\\lambda _3^{\chi \Delta }&=\lambda _7^{\chi \Delta }=0.
\end{align}
\end{document}